# Big Data Intelligence Using Distributed Deep Neural Networks

Felix Ongati
University of Nairobi
School of Computing and Informatics
Email: felixongati@gmail.com

Dr. Eng. Lawrence Muchemi
University of Nairobi
School of Computing and Informatics
Email: lmuchemi@uonbi.ac.ke

*Abstract*

Large amount of data is often required to train and deploy useful machine learning models in industry. Smaller enterprises do not have the luxury of accessing enough data for machine learning, For privacy sensitive fields such as banking, insurance and healthcare, aggregating data to a data warehouse poses a challenge of data security and limited computational resources. These challenges are critical when developing machine learning algorithms in industry. Several attempts have been made to address the above challenges by using distributed learning techniques such as federated learning over disparate data stores in order to circumvent the need for centralised data aggregation.

This paper proposes an improved algorithm to securely train deep neural networks over several data sources in a distributed way, in order to eliminate the need to centrally aggregate the data and the need to share the data thus preserving privacy. The proposed method allows training of deep neural networks using data from multiple de-linked nodes in a distributed environment and to secure the representation shared during training. Only a representation of the trained models (network architecture and weights) are shared.

The algorithm was evaluated on existing healthcare patients data and the performance of this implementation was compared to that of a regular deep neural network trained on a single centralised architecture. This algorithm will pave a way for distributed training of neural networks on privacy sensitive applications where raw data may not be shared directly or centrally aggregating this data in a data warehouse is not feasible.

## I. Introduction

Enormous amount of data is required for useful application of machine learning techniques in industry. In the recent past, most big tech companies such as Google, Facebook and Amazon have been aggregating data from multiple sources, joining it together and applying artificial intelligence techniques on this data to generate useful business insights. This gives the data aggregators an unfair advantage over their small competitors who do not have enough infrastructure for big data aggregation and analysis.

In domains such as healthcare(EMR and EHR) and finance (banking and credit card transactions), large volumes of data is collected daily (including unstructured data), which poses a challenge to the computational resources available to centrally aggregate this data in a data warehouse. Raw data cannot be shared directly due to the sensitive nature of the data, and this data is not necessarily labelled or structured. Most large companies in these domains decentralize their information and transaction management systems or adopt horizontal scaling, so as to secure their clients data and optimize utilization of scarce computing resources.

Adoption of artificial intelligence techniques in business to enhance the quality of products, optimize planning, maximize output and improve customer satisfaction is at an all time high. Modern companies are now moving from mere predictive analytics to prescriptive analytics and possibilities beyond. We are now talking of artificial super intelligence (ASI), as path towards achieving technological singularity. It is only a matter of time for the debate on the achievability of technological singularity as to surpass human intelligence to be settled, but right now we cannot overlook the possibilities that have been brought about by artificial intelligence, especially in enterprise operations. Data sensitive operations such as targeted marketing, that were traditionally based on purely human efforts have experienced great positive results attributed to adoption of AI techniques. To reap the full potential benefits of using artificial intelligence, enough data must be available to build models and extract knowledge. Labelled customer data cannot be shared for this purpose due to privacy issues and sensitivity of customer information that can land in unauthorised hands from such data.

For small and medium enterprises, accumulating enough data for the application of machine learning techniques to optimize business processes is almost unrealistic due to the time needed to have substantial data for such purposes.

Successfully deployment of artificial intelligence solutions requires development of a very deliberate internal data sharing culture. Siloed data is one of the greatest killers of AI application in industry (Medium, 2018). Since machine learning problems are very data intensive, corporations wishing to explore machine learning have to first breakdown data silos, not only to increase the amount of data on which an AI model can be trained, but also to increase the diversity of the data. Though breakdown the data silos sounds like a brilliant idea, it becomes challenging in two ways;

1. When this data is individual customers' data and breaking down the silos to share it will infringe on their privacy.
2. When resources to gather centrally and store data from these diverse sources are limited.

From challenge (2) above, the following pertinent question arises; how can health facilities, insurance companies and other SMEs with smaller or more specialized datasets benefit given that huge and diverse datasets are required to train and deploy usable modern machine learning systems? Can small and medium enterprises collaboratively train a common model without sharing data stored in their databases? Thanks to recent progress in deep learning research, it is now possible for all corporations to benefit through sharing learnt representations other that sharing data. It is now possible to work with unlabelled and unstructured big data using deep neural networks and autoencoders. The success of machine learning methods in extracting patterns from structured data has led to its adoption in processing unstructured data such as images. In deep learning, the focus is on learning representation apart from learning patterns. Advancement in representational learning makes it possible for a trained model to share its knowledge with other deep learning systems.

This project explores solutions to the above questions by modeling and simulating distributed learning to train a common model over disparate databases consisting of medical xray images.

**Sharing Representation Instead of Sharing Data**

Since it is possible to share learned representation between different deep learning systems instead of sharing raw data, data security and privacy will no longer be a major issue in AI adoption. A description of the data, description of training model and the learned weights is what will need to be shared with other learning systems that would need to run the same on their data. Deep learning provides opportunities for continuous learning and improvement for AI systems as they get exposed to different data sets. As Philippe Beaudoin (Medium, 2018) correctly points out; advances in representation learning will to a great extend enable sharing of expertise and in the future, the greatest killer of AI was siloed expertise.

II. WORKS OF INTEREST

**Federated Learning at Google AI Lab (Google AI Blog, 2018)**

Federated learning is an approach to training models from user interaction with mobile devices; models are trained on the devices. Federating learning decouples machine learning from data storage by enabling learning a shared prediction model with the training data on the device without sharing the data (McMahan and Ramage, 2017). In this approach, there is a shared model that resides in the cloud.

A mobile device gets the current up-to-date shared model from the cloud, runs it on the local data on the phone and stores the improvements as a small focused update. These changes to the model are then sent to the cloud, averaged with updates from other devices and applied to the central main model in the cloud. No training data leaves the users' devices.

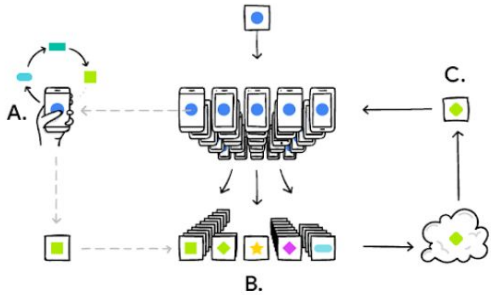

*Figure 1: Federated learning architecture*

A mobile device (A) localizes the model in the context of a user's interaction with the device. Users' changes are summed up (B) to form an aggregate change (C) which is then applied to the shared model and procedure repeated whenever new data is available. (Google AI Blog, 2018). Federated learning proposes a mechanism suitable for training centralized models in an unreliable network connection environment where sharing data would be expensive in addition to privacy concerns. This work borrows a lot from federated learning techniques, albeit on training a decentralised model.

**Distributed Learning of Deep Neural Network Over Multiple Agents (Gupta and Raskar, 2018)**
This work by Gupta and Raskar explores an algorithm that allows for distributed deep learning over multiple agents. The algorithm was evaluated on existing mixed NIST (MNIST) datasets and it was shown that the obtained performance was similar to that of a regular neural network trained on a centralised data source. Gupta and Rasker raised security concerns on their algorithm evaluation which we seek to address here by running an improved algorithm on privacy sensitive electronic medical records patients' data.

**Blockchain Technology**
A blockchain is an open distributed, decentralized, public ledger that securely stores transaction records between parties by using public key cryptography on a peer-to-peer network (Harvard Business Review, 2019).

Transactions are recorded as a block of data which is hashed and distributed across multiple nodes in the network. Each block is linked to a preceding block in their order of generation, forming a logical chain.

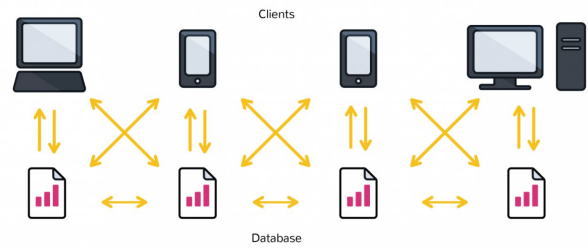

*Figure 2.4: High level blockchain decentralization concept (CoinDesk, 2019)*

Each block is generated by application of a complex mathematical computation that is resource intensive. Coupling this generation process with public key cryptography and the fact that the same block is replicated to many nodes in the network makes blockchain technology inherently secure.

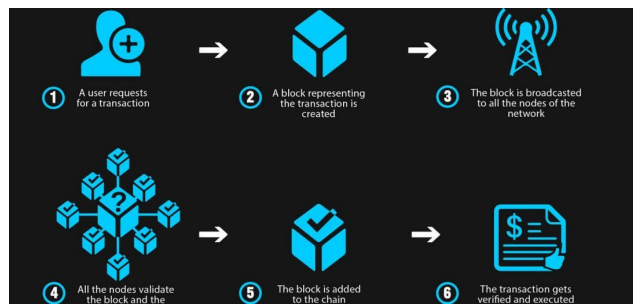

*Figure 2.5: Simplified chain generation process (Anwar, 2019)*

Blockchain is not completely anonymous but confidential, transactions are publicly recorded on the blockchain without complete user data. A user's private key is created by shortening the public key through a complex mathematical algorithm. Security of blockchain technology is pegged on the fact that it is practically impossible to reverse the process and generate a private key from a public key due to the complexity of the algorithm used. This is how blockchain technology achieves its confidentiality.

Implementation of the algorithm used in this research project borrows a lot from blockchain technology with few modifications but totally different objectives. They both run on distributed network, use private key cryptography and every node on the network comes to the same conclusion, each updating their records independently. Blockchain technology requires authorization and authentication to establish trust whereas in our case, the keys are used for encryption only to conceal weight transfer between nodes and not for authentication/authorization.

### III. METHODOLOGY

To answer each of the questions, this research involved theoretical studies of existing literature and implemented projects, experimental studies, simulation, prototyping and empirical evaluation.

**Theoretical Studies**

To find out an optimal algorithm for securely training a distributed deep learning model in a distributed big data environment, there was a review of existing literature. This involved reviewing of academic papers and white papers in the area of distributed machine learning on big data and centralised machine learning on big data environments. A study of other distributed machine learning open source and commercial projects was also conducted, including a study of federated learning approach widely used on training centralised models in mobile devices such as smartphones and tablets. The output of this study was an optimal algorithm that can be used to securely train a distributed deep learning neural network model in a distributed big data environment. Arriving at an optimal algorithm required coming up with a hybrid algorithm that borrows the best techniques from each of the identified approaches.

A study of the design of distributed machine learning open source projects was useful in designing a simulation of a distributed environment.

**Simulation and Prototyping**

This paper prototypes a distributed deep learning environment using python parallel programming and secure shell protocol (SSH) for file transfer and communication; sending model architecture and weights. Convolution neural network was used for automatic feature extraction and representational learning on the data. The environment was setup on a personal computer with Nvidia GTX 1050 Graphics Processing Unit, running on Ubuntu 18.04 operating system with Python 3.6 with installed with Numpy, Tensorflow and Keras deep learning libraries.

**Empirical Evaluation**

The number of nodes in the distributed environment was increased periodically as performance was being monitored in terms of time taken to converge and accuracy so as to generate concrete recommendations for future work.

The performance of this model of training a distributed deep neural network on big data was compared against implementing the same model on a centralised deep learning environment where all the data is centrally aggregated. The comparison was in terms of accuracy of the clustering and time taken for the models to converge. Holdout validation was used to evaluate the accuracy of the model by itself, where the data was split at a ratio of 80/20, 80 percentage of each class for training and 20 percent of each class for testing.

### IV. DATASET AND ALGORITHM IMPLEMENTATION

**Dataset**

The data used to train the model and test the algorithm consisted of chest x-ray images obtained from 28780 patients. There were two sets of data; xray images from normal patients and those from patients with pneumonia.

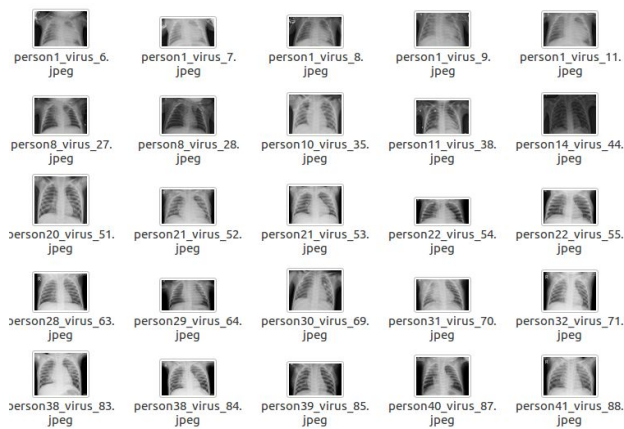

Figure 2: Dataset

The binary images were reformatted and spatially normalized to fit in a 20 × 20 bounding box. The images were manually labelled by adding a prefix of 1 to the file names of those from patients with pneumonia and prefix 0 to file names of those from patients without pneumonia. The images were then loaded to a Numpy array and the sizes scaled by a factor of 1/255 to reduce space and time complexity.

**Algorithm Design**

This algorithm implements a technique that was used to train deep neural networks over multiple data sources in a secure electronic medical records environment while mitigating the need to share raw data directly. The algorithm implementation used three categories of agents as below;

I. Alice → deep neural network that runs at the data source nodes.

II. Bob → acts as a watchdog, monitoring file changes (model improvements) which are encrypted and send to Charity. Bob also gets initial models from Charity when Alice wants to train on new data.
III. Charity → remote deep neural network that coordinates sharing of weights among the training nodes and stores up-to-date model architecture.

**START**
I. Charity initializes a network architecture with random weights and other parameters.
II. Charity compiles the model and trains with its local data for 1 epoch, writing the best output model to a file.
III. Charity encrypts the model file using her own private key.
IV. When $Alice_k$ wants to train;
    a. $Bob_k$ fetches the current model from Charity via secure shell (SSH) protocol
    b. $Bob_k$ decrypt the model file using Charity's public key and makes it available for $Alice_k$
V. $Alice_k$ reads the model, compiles it trains (a series of forward and back propagation) the model using its local data.
VI. While training, $Alice_k$ monitors model improvements in terms of accuracy and loss at the end of every epoch, with an early stopping patience of 10.
VII. If there is an improvement, $Alice_k$ saves the improved model to a file, this alerts $Bob_k$.
VIII. $Bob_k$ encrypts the new improved model file using Charity's public key and send the file to Charity
IX. Upon receipt, Charity runs the model from $Bob_k$ on local data while comparing with her current model
X. If the new model from $Bob_k$ is better than the current model, Charity saves the new model as the currently available model, or else the model from $Bob_k$ is discarded.
XI. When $Alice_{k+1}$ wants to train, it will use the currently available model from Charity as the starting point.
XII. End

**Pictorial Representation of the Algorithm**

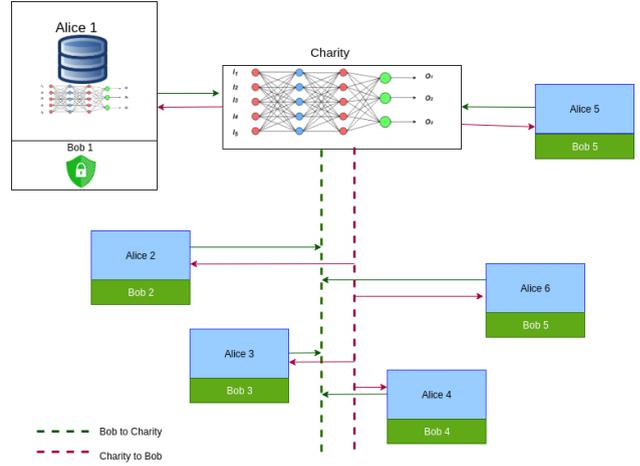

Figure 3: Pictorial representation of the algorithm.

## V. RESULTS AND ANALYSIS

### 1. Accuracy

Accuracy was used to evaluate the performance of this model. The plots below indicate that the model generally improved every epoch until it hit a maximum accuracy of 93.63% before smoothing and and a final accuracy of 93.55% after smoothing. Accuracy was calculated by finding the ratio between the correctly predicted classes and the total number of predictions. The slope of this curve is relatively similar to one achieve by training a centralised model on a centralised database.

**Accuracy before smoothing**
Maximum Value: 99.81%

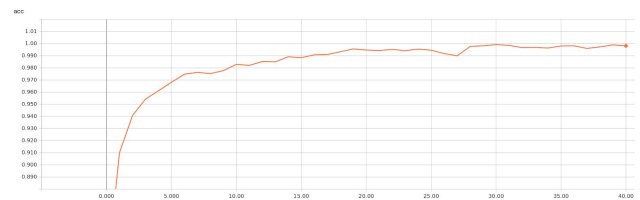

Figure 4: Accuracy before smoothing; y-axis = accuracy, x-axis = number of epochs

**Accuracy after smoothing**
Initial Value: 0
Final Value: 99.81%
Smoothing Factor: 0.6

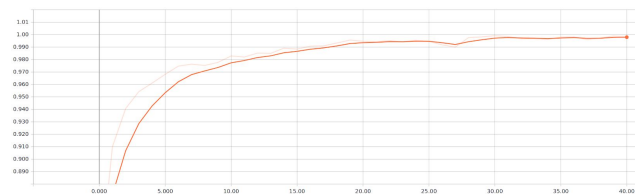

Figure 5: Accuracy after smoothing; y-axis = accuracy, x-axis = number of epochs

### 2. Loss

The objective of the model was to minimise loss while increasing accuracy. During each iteration of training, the loss was calculated and used to evaluate current performance, and tweak the model parameters based on this feedback. Binary cross entropy was used as the loss function. From the graph below, the loss decreases steadily in every epoch.

**Loss before smoothing**

Minimum Value: 0.1936

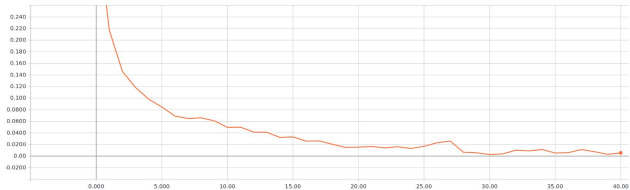

Figure 6: Loss before smoothing; y-axis = loss, x-axis = number of epochs

**Loss after smoothing**

Final Value: 5.4970e-3

Smoothing Factor: 0.6

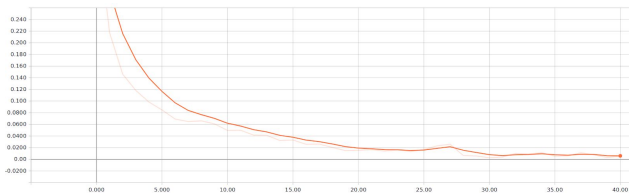

Figure 7: Loss after smoothing; y-axis = loss, x-axis = number of epochs

### 3. Comparison with centralised models

The same model was also trained on centralised architecture. There was no difference between the distributed architecture and the centralised architecture in terms of final loss and accuracy achieved, but there was a significant difference in terms of time taken to converge. Each node on the distributed network training on about 37000 images, doing 100 epochs and a batch size of 200 images took an average of 4 minutes to converge. For a centralised architecture with the same number of images, batch size, epoch and same infrastructure resource configuration as one node on the distributed architecture, took close to 3 times average time taken on the distributed learning architecture.

### 5. Effect of increasing the number of layers on accuracy and loss

There was no significant difference on accuracy and loss when the number of hidden layers was doubled from 4 to 8 but the was a significant different on the number of epochs required and time taken to converge as shown on the charts below;

**Accuracy**

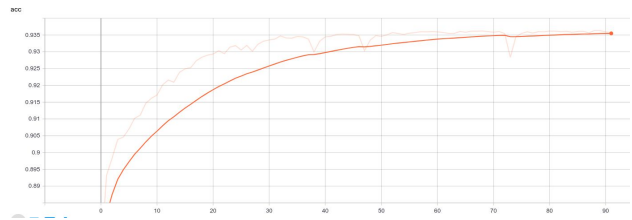

Figure 8: Accuracy after doubling number of hidden layers; y-axis = accuracy, x-axis = number of epochs

**Loss**

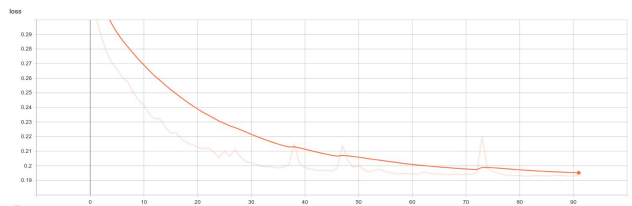

Figure 9: Accuracy after doubling number of hidden layers; y-axis = accuracy, x-axis = number of epochs

## VI. CONCLUSION AND FUTURE WORK

**Conclusion**

This research work has proved beyond reasonable doubt the practicability of training a shared deep learning model in an unstructured data environments without having to share actual data. Eliminating the need to share data or the need to aggregate data in a data warehouse ensures security of the data in the nodes where it is generated. Communication security while exchanging weights and other network parameters can be ensured by leveraging private-public authentication on SCP,SFTP and SSH mechanisms. Thus databases that handle sensitive data such as patients' medical records in electronic medical records can benefit from the application of machine learning techniques on their data without compromising on data security and privacy of patients.

This approach can also be beneficial in low resource and low data scenarios by training a deep learning model on several smaller databases as if they were a single data repository. Thus big data, data security and limited computational resources should not hinder developing and adopting machine learning algorithms in industry.

**Recommendations and Future work**
Though this approach has been proven to work here, having a centralised coordinator (Charity) will be will break the whole network whenever Charity is down or his resources are overwhelmed by the number of nodes connecting. A reasonable extension to this research work can be to develop a technique to have several coordinators or who are synchronised and the nodes can connect to the nearest one. An algorithm can also be developed to have the nodes intelligently elect a new coordinator whenever the central coordinator breaks down. It would also be interesting to explore using blockchain technology to shared model parameters in situations where each of the nodes should remain completely hidden from the other nodes or where each node should be rewarded based on their contribution to the final model.

References

1. Blueshift.com. (2018). *The Leading Customer Data and Engagement Platform for Marketers*. [online] Available at: https://blueshift.com [Accessed 24 Oct. 2018].
2. Columbus, L. (2018). *77% Of Marketing Execs See AI Adoption Growing This Year*. [online] Forbes. Available at: https://www.forbes.com/sites/louiscolumbus/2018/05/14/77-of-marketing-execs-see-ai-adoption-growing-this-year/#45af2fd77ef8 [Accessed 24 Oct. 2018].
3. Medium. (2018). *The future of AI: from siloed data to shared expertise*. [online] Available at: https://medium.com/element-ai/the-future-of-ai-from-siloed-data-to-shared-expertise-a5391ee18444 [Accessed 24 Oct. 2018].
4. Decentralizedml.com. (2018). *Decentralized Machine Learning - DML*. [online] Available at: https://decentralizedml.com/ [Accessed 25 Oct. 2018].
5. BigAI. (2018). *Blockchain based Decentralized Artificial Intelligence & Machine Learning - BigAI*. [online] Available at: https://bigai.io/en/ [Accessed 25 Oct. 2018].
6. Chervenak, A., Foster, I., Kesselman, C., Salisbury, C. and Tuecke, S. (2000). The data grid: Towards an architecture for the distributed management and analysis of large scientific datasets. *Journal of Network and Computer Applications*, 23(3), pp.187-200.
7. Snips. (2018). *Snips — Using Voice to Make Technology Disappear*. [online] Available at: https://snips.ai/ [Accessed 27 Nov. 2018].
8. McMahan, B. and Ramage, D. (2017). *Federated Learning: Collaborative Machine Learning without Centralized Training Data*. [online] Google AI Blog. Available at: https://ai.googleblog.com/2017/04/federated-learning-collaborative.html [Accessed 28 Nov. 2018].
9. McMahan, B., Moore, E., Ramage, D., Hampson, S. and Aguera, B. (2017). Communication-Efficient Learning of Deep Networks from Decentralized Data.
10. Bonawitz, K., Ivanov, V., Kreuter, B., Marcedone, A., McMahan, B. and Patel, S. (2017). Practical Secure Aggregation for Privacy-Preserving Machine Learning.
11. Openmined.org. (2018). *OpenMined*. [online] Available at: https://openmined.org/ [Accessed 3 Dec. 2018].
12. Rodriguez, J. (2018). *Technology Fridays: OpenMined Powers Federated AI Using the Blockchain*. [online] Towards Data Science. Available at: https://towardsdatascience.com/technology-fridays-openmined-powers-federated-ai-using-the-blockchain-d124e6560dd [Accessed 3 Dec. 2018].
13. Cohen, G., Afshar, S., Tapson, J. and van Schaik, A. (2017). EMNIST: an extension of MNIST to handwritten letters.
14. IBM.com. (2018). *Big Data Analytics*. [online] Available at: https://www.ibm.com/analytics/hadoop/big-data-analytics [Accessed 6 Dec. 2018].
15. Bernes, J. (2017). *The AI-First Business Model – Element AI – Medium*. [online] Medium. Available at: https://medium.com/element-ai/the-ai-first-business-model-fcc41c069440 [Accessed 6 Dec. 2018].
16. Gupta, O. and Raskar, R. (2018). Distributed learning of deep neural network over multiple agents. *Journal of Network and Computer Applications*, 116, pp.1-8.